arXiv:1203.0702# Electronic structure and enhanced visible light absorption of N, B-codoped TiO$_2$

**Meili Guo** [1,2], **Xiao-Dong Zhang**[1], **and Jiulin Du**[*,1]

[1] Department of Physics, School of Science, Tianjin University, No.92, Weijin Road, Tianjin 300072, People's Republic of China
[2] Department of Physics, Tianjin Institute of Urban Construction, No.26, Jinjing Road, Tianjin 300384, People's Republic of China**Keywords** (TiO$_2$, Codoping, Electronic structure, Visible absorption)

* E-mail: jiulindu@yahoo.com.cn**Abstract** We present the GGA+U calculations to investigate the electronic structure and visible light absorption of the N, B-codoped anatase TiO$_2$. The N$_s$B$_i$ (substitutional N, interstitial B) codoped TiO$_2$ produces significant Ti 3$d$ and N 2$p$ mid-gap states when the distance of N and B atoms is far, and the N$_i$B$_i$ (interstitial N and B) and N$_s$B$_s$ (substitutional N and B) codoped TiO$_2$ prefer to form localized $p$ states at 0.3-1.2 eV above the valence band maximum. Further, the optical band edges of the three codoped systems shift slightly to the visible region, but only the far distance N$_s$B$_i$ codoped TiO$_2$ shows an obvious visible optical transition. These results indicate that the N$_s$B$_i$ codoped TiO$_2$ has a dominant contribution to the visible absorption of the N, B-codoped TiO$_2$.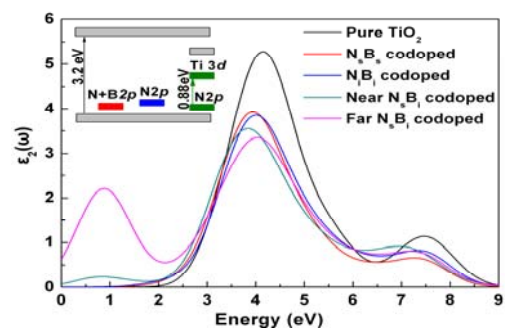

First-principles GGA+U calculation reveals that the far distance N$_s$B$_i$ (substitutional N and interstitial B) doping can cause the N 2$p$-Ti 3$d$ transition and can obviously enhance the visible optical absorption of TiO$_2$.

Due to having strongly localized N 2$p$ states (0.3-0.5 eV) at the top of valence band, the N-doped TiO$_2$ is very promising to enhance the optical absorption in the visible region and improve the visible photoactivity.[1] Meanwhile, the mid-gap states of B atom are found in the B-doped TiO$_2$.[2,3] In the previous work, it was found that the interstitial B doped in TiO$_2$ could shift the band edge to the high energy region, and only increase the photocatalytic activity under the UV light irradiation by forming Ti-O-B.[4] However, Lu *et al.* reported that the Ti−B−O structure in the B-doped TiO$_2$ could decrease the optical band gap from 3.34 to 3.1 eV, exhibiting a higher photoconversion efficiency under the UV and 400-620 nm visible light irradiation.[5] Subsequently, the standard density functional theory (DFT) calculation showed that the substituted B in TiO$_2$ formed the mid-gap states at 1.5 eV above the valence band maximum (VBM).[3] Recently, the experimental nuclear magnetic resonance (NMR) and electron paramagnetic resonance (EPR) showed the results that the Ti$^{3+}$ species were the dominant states in the B, F-codoped TiO$_2$, which induced some mid-gap states in the gap and improved the visible photoactivity.[6] Based on such ideas, the N, B-codoped TiO$_2$ has been proposed as the visible light photocatalyst. In 2007, In *et al.* reported that as compared with the B or N single doped system, the N, B-codoped $TiO_2$ could improve the visible absorption and the photocatalytic activity.[7] And subsequently, it received the support by the EPR experiments, which conceived that the B and N atoms could substitute the nearest neighbor O and then form $N_sB_s$ structure (substitutional N and B).[8] However, the recent NMR and EPR data demonstrated clearly that another novel $N_iB_i$ structure (interstitial N and B) played an important role in the N, B-codoped $TiO_2$ system, and this was the main reason for the visible light photoactivity.[9, 10] Moreover, Feng et al.[9] observed two kinds of interstitial B in the N, B-codoped $TiO_2$, which was in good agreement with the hybrid functional calculations.[11] In the light of present inconsistency in the N, B-codoped $TiO_2$, we try to reveal the physical nature of these structures and their optical transition mechanism. In this letter, we investigated three kinds of structures of the N, B-codoped $TiO_2$ by using the GGA+U corrections, namely $N_sB_s$ proposed by Gopal,[8] $N_sB_i$ (substitutional N, interstitial B), and $N_iB_i$ proposed by Czoska [See Fig.1].[10]

In our calculations, CASTEP was used, which was based on the DFT using the plane-wave pseudopotential method.[12] We used the generalized gradient approximation (GGA) in the scheme of Perdew-Burke-Ernzerhof (PBE) to describe the exchange-correlation functional.[13] The ultrasoft pseudopotential was used to describe the electron-ion interaction.[14] In our GGA+U calculations, the on-site effective U parameter (3.5 eV) was applied to the Ti $3d$ states. Then, the band gap of the pure $TiO_2$ could be corrected to 3.12 eV compared with 2.18 eV of the pure GGA. very close to the experiment value of 3.2 eV. Two anatase $TiO_2$ supercells were chosen, containing 72 and 36 atoms respectively, to calculate the electronic structures at different concentrations. We chose the energy cutoff as 380 eV. And the Brillouin-zone sampling mesh parameters for the $k$-point set were 2×1×2 and 4×1×2, corresponding to the 72 and 36 atoms systems respectively.[15] The charge densities were converged to $2 \times 10^{-6}$ eV/atom in the self-consistent calculation. In the optimization process, the energy change, the maximum force, the maximum stress, and the maximum displacement tolerances were set $2 \times 10^{-5}$ eV/atom, 0.05 eV/Å, 0.1 Gpa, and 0.002 Å, respectively.

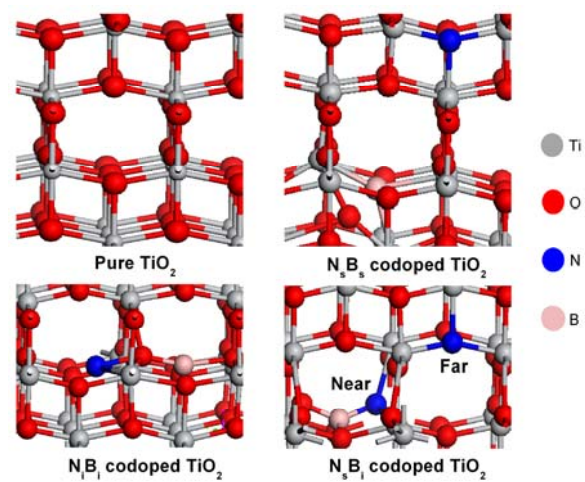

**Figure 1** Optimized local structures of the pure, $N_sB_s$, $N_iB_i$, and $N_sB_i$ codoped anatase $TiO_2$, and the supercell containing 72 atoms. Two structures, namely near and far configuration, are considered for the $N_sB_i$ codoped anatase $TiO_2$.

Fig.2 (a) gives the total density of states (TDOS) of the pure, $N_sB_s$, $N_iB_i$, and $N_sB_i$ codoped $TiO_2$. Compared with the band gap of 3.12 eV for the pure $TiO_2$, the band gaps of the $N_sB_s$, $N_iB_i$, and near distance and far distance $N_sB_i$ codoped $TiO_2$ decrease to 2.98, 3.10, 2.80, and 2.88 eV, respectively. The doping states are observed in the band gap for three codoped structures [See Fig.2 (b)]. For the $N_sB_s$ codoped structure, new states are observed at 0.68 eV above the VBM, which are mainly composed of the N and B $2p$ states. Meanwhile, the energy levels with 1.22 eV appear above the VBM of the $N_iB_i$ codoped $TiO_2$, but these electronic states are mainly due to the N $2p$ and Ti $3d$ states, while the B $2p$ states have a contribution which is negligible by analyzing the corresponding partial density of states (PDOS) in Fig.2 (b). The near distance $N_sB_i$ codoping causes the electronic states with 0.91 eV above the VBM, which are mainly due to the Ti 3*d* states. The far distance $N_sB_i$ codoping causes the energy levels with 0.45, 1.31, and 2.65 eV above the VBM, corresponding to the N 2*p*, Ti 3*d*, and Ti 3*d* states, respectively. It is worth noting that, in the far distance $N_sB_i$ codoped $TiO_2$, the B 2*p* sates still have a contribution negligible to the doping electronic states in the gap. In the three codoped systems, the N 2*p* states have a dominant effect on the doping states. The Ti atom can cooperate with the adjacent doping atom and cause the Ti 3*d* states shift to the top of valence band in the $N_sB_s$ and $N_iB_i$ codoped structures and the bottom of conduction band in the $N_sB_i$ codoped structure.[16] In addition, the significant difference between spin-up and spin-dawn states indicates that the $N_sB_s$ and $N_sB_i$ structures include lots of unpaired electrons, which are in good agreement with the EPR results.[9, 10] We also calculated the electronic structures of 36 atoms systems and find that the band gap is very similar to that of 72 atoms systems, although the doping states have a slight shift.

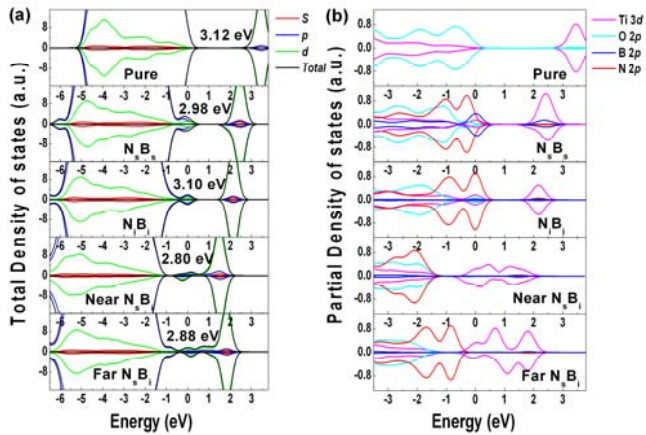

**Figure 2** The TDOS of the pure, $N_sB_s$, $N_iB_i$, and near distance and far distance $N_sB_i$ codoped $TiO_2$ in 72 atoms systems, and the *s*, *p*, *d* states are sums over all atoms (a), and the corresponding PDOS for Ti 3*d*, O2*p*, N2*p*, and B2*p* (b).

Fig.3 shows the imaginary part of dielectric function $\varepsilon_2(\omega)$ (a) and the optical absorption (b) of the pure, $N_sB_s$, $N_iB_i$, and $N_sB_i$ codoped $TiO_2$. For the pure $TiO_2$, the optical transition of 4.13 eV ($E_g$) is only observed in Fig.3 (a), which is ascribed to the intrinsic transition between the O 2*p* states in the highest valence band and the Ti 3*d* states in the lowest conduction band. After the N and B incorporation, the optical transition of $E_g$ decreases to 3.93, 4.03, 3.84, and 4.05 eV for the $N_sB_s$, $N_iB_i$, and near distance and far distance $N_sB_i$ codoped $TiO_2$, respectively, which means that the band gaps of the doping systems are narrowed. Meanwhile, the visible optical transition ($E_1$) at 0.88 eV is observed in the near distance and far distance $N_sB_i$ codoped $TiO_2$. In Fig.2, the electronic structure calculation clearly shows that the occupied N 2*p* states locate at 0.45 eV, and the unoccupied Ti 3*d* states appear at 1.31 eV above the VBM in the far distance $N_sB_i$ codoped $TiO_2$, indicating that the optical transition of $E_1$ is due to the N 2*p*-Ti 3*d* transition. Additionally, in the near distance $N_sB_i$ codoped $TiO_2$, the intensity of the $E_1$ is relatively low. Fig.3 (b) presents the optical absorption of the pure and three codoped systems. It is shown that the optical band edges of the $N_sB_s$, $N_iB_i$, and $N_sB_i$ codoped $TiO_2$ shift slightly to the visible light region, and the optical band gaps decrease from 3.27 eV for the pure $TiO_2$ to 3.05, 2.98, 2.75, and 3.0 eV respectively for the $N_sB_s$, $N_iB_i$, and near distance and far distance $N_sB_i$ codoped $TiO_2$, which are in good agreement with the recent experimental data.[9] The decreased band gap shows that the localized N and B 2*p* electronic states both participate in the shift of band edge in the $N_sB_s$ codoped $TiO_2$, but the N 2*p* electronic states have a dominant contribution to the band edge shift in the $N_iB_i$ and $N_sB_i$ codoped $TiO_2$. The visible absorption has been observed in the $N_sB_i$ codoped $TiO_2$, which suggests that the near infrared optical absorption observed experimentally is due to the cooperation interaction between substitutional N and interstitial B in $TiO_2$.[17] Furthermore, we find that the $N_sB_i$ structure is sensitive to the distance, and a decrease of the distance between N and B atoms induces a decrease of absorption. We also consider the effect of dopants positions on the $N_sB_s$ and

$N_iB_i$ codoped $TiO_2$, and find that the optical properties have not been influenced by the distance of doping atoms.

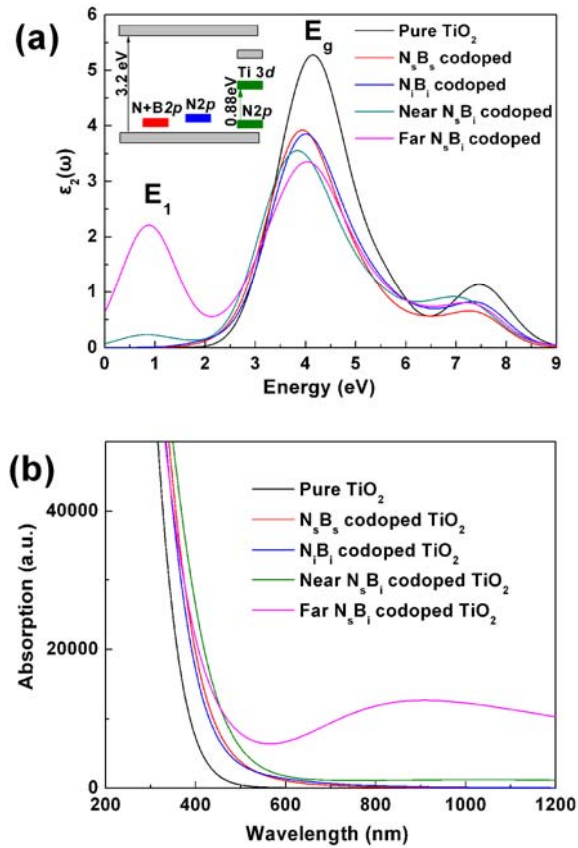

**Figure 3** The imaginary part of dielectric function $\varepsilon_2(\omega)$ (a) and the optical absorption (b) of the pure, $N_sB_s$, $N_iB_i$, and near distance and far distance $N_sB_i$ codoped $TiO_2$ in 72 atoms systems. Inset is an outline of the optical transitions of the $N_sB_i$ codoped $TiO_2$.

In summary, we have evaluated the electronic and optical properties of the N, B-codoped $TiO_2$ by the DFT method. The double substitutional ($N_sB_s$) doping leads to an enhancement of $p$ states above the VBM and produces an optical band edge shift. The N and B electronic states both have a significant contribution to the doping states. The double interstitial ($N_iB_i$) doping can also cause mid-gap states, but only the N $2p$ states have a contribution to the optical band edge shift. The substitutional N and interstitial B ($N_sB_i$) doping can cause the mid-gap states not only, but also can introduce the obvious visible absorption which is attributed to the N $2p$-Ti $3d$ transition.

**Acknowledgements** This work is supported by National Natural Science Foundation of China (Grant No. 11175128) and Higher School Specialized Research Fund for Doctoral Program (Grant No. 20110032110058).